\documentclass{PoS}

\usepackage{amsfonts}
\usepackage{graphicx}
\usepackage{amsmath}
\usepackage{bm}

\newcommand*{\chpt}{\raise0.4ex\hbox{$\chi$}PT}

\newcommand{\Delstar}{\ensuremath{\Delta^{\raise0.18ex\hbox{${\scriptstyle *}$}}}}
\def\gtwid{{\,\raise.35ex\hbox{$>$\kern-.75em\lower1ex\hbox{$\sim$}}\,}}
\def\ltwid{{\,\raise.35ex\hbox{$<$\kern-.75em\lower1ex\hbox{$\sim$}}\,}}
\def\leftvec{{\raise1.5ex\hbox{$\leftarrow$}\kern-1.00em}}
\def\rightvec{{\raise1.5ex\hbox{$\rightarrow$}\kern-1.00em}}
\def\half{{\scriptstyle \raise.2ex\hbox{${1\over2}$}}}
\def\threehalves{{\scriptstyle \raise.15ex\hbox{${3\over2}$}}}
\def\third{{\scriptstyle \raise.15ex\hbox{${1\over3}$}}}
\def\third{{\scriptstyle \raise.15ex\hbox{${1\over3}$}}}
\def\twothirds{{\scriptstyle \raise.15ex\hbox{${2\over3}$}}}
\def\fourth{{\scriptstyle \raise.15ex\hbox{${1\over4}$}}}

\newcommand*{\bea}{\begin{eqnarray}}
\newcommand*{\eea}{\end{eqnarray}}
\newcommand*{\be}{\begin{equation}}
\newcommand*{\ee}{\end{equation}}

\title{Finite Volume Study of the Delta Magnetic Moments Using Dynamical Clover Fermions}

\ShortTitle{Finite Volume Study of the Delta Magnetic Moments}

\author{\speaker{C.\ Aubin}\\
        Department of Physics, College of William and Mary, 
        Williamsburg, VA, USA\\
        E-mail: \email{caaubin@wm.edu}}

\author{K.\ Orginos\\
        Department of Physics, College of William and Mary, 
        Williamsburg, VA, USA\\
        Jefferson Laboratory, Newport News, VA, USA}

\author{V.\ Pascalutsa and M.\ Vanderhaeghen\\
        Institut f\"ur Kernphysik, Universit\"at Mainz, 
        Mainz, Germany\\
        }

\abstract{We calculate the magnetic dipole moment of the $\Delta$ baryon using a background magnetic field on 2+1-flavors of clover fermions on anisotropic lattices. We focus on the finite volume effects that can be significant in background field studies, and thus we use two different spatial volumes in addition to several quark masses.}

\FullConference{The XXVI International Symposium on Lattice Field Theory\\
		 July 14-19 2008\\
		 Williamsburg, Virginia, USA}

\begin{document}

\bibliographystyle{JHEP}

\section{Introduction}

In order to calculate electromagnetic properties of baryons, such as the magnetic dipole moment, one has two options on the lattice. One method is to calculate the electromagnetic form factors, which involves a three-point function calculation as well as a difficult extrapolation to the $q^2=0$ point. The other method is to implement a classical background field, measure the two-point function and extract the magnetic moment from the $B$-field dependent mass (first done in Ref.~\cite{Bernard:1982yu}, and most recently in Ref.~\cite{Lee:2005ds}\footnote{See references within Ref.~\cite{Lee:2005ds} for other calculations of baryon magnetic moments with this technique.}). This latter method is much cleaner and simpler if we are only interested in the static moments, and this is the approach we shall take here. 

It is rather straightforward to implement a background field in a lattice simulation. On a given configuration, we multiply all of the $SU(3)$ gauge fields by a $U(1)$ gauge field, given by
\be
	U_\mu(x) = \exp\left[iq a A_\mu(x)\right] \ ,
\ee
where $q$ is the charge of the quark whose propagator we are calculating. For a constant magnetic field with a magnitude of $B$ pointing in the $+z$-direction, the traditional choice is 
\be\label{eq:y-link-mod}
	A_\mu(x,y,z,t) = a B x\; \delta_{\mu y}\ .
\ee
Thus, all of the $y$-links are modified by $\exp\left[i q a^2 B x\right]$, and all other links are unchanged. With this \cite{Bernard:1982yu}, one can calculate a baryon two-point function which behaves for large time in the usual manner
\be\label{eq:2pt}
	C_{s,B}(t) \sim A_{s,B}\ e^{-m_{s,B} t} + \ldots\ ,
\ee
but with the exponential damping governed by a $B$-field dependent mass
\be\label{eq:mB}
	m_{s,B} = m_0 \pm s\mu B + O(B^2)\ .
\ee
Here, $m_0$ is the mass of the baryon in the absence of any external field. The magnetic moment is given by
\be\label{eq:magmom}
	\mu_z = g\frac{e}{2m_0}S \ ,
\ee
where we take the spin to be in the $z$-direction, and $S$ is the total spin. The variable $s$ in Eq.~(\ref{eq:mB}) equals $S_z/S$, and for $S=3/2$, we can have $s=\pm1/3, \pm1$. This is useful when one calculates all possible spin projections of a given baryon, which we do for all of our results in this work. The plus (minus) sign is when the spin is anti-parallel (parallel) to the magnetic field, and we are ignoring both excited states and non-linear (in $B$) terms.

On a technical note, the number input into the simulation is $qa^2B$, and thus includes the product of the quark charge and the magnetic field in lattice units. In order to account for the quark charges of the up, down, and strange quarks, for a single magnetic field $B$ we must use two values of $q a^2 B$, corresponding to the fact that $q_u = -2q_{d,s}$. For particles made up with only a single quark ($\Delta^{++},\Omega^-$), we need one input value, but for the $\Delta^+$ or the nucleon, for example, we must use two inputs that differ by factors of -2 so the quarks all experience the same $B$-field. 

\section{Periodicity constraints}

Given that simulations are done on a finite volume, there is an immediate consequence when we hit the boundary in the $x$-direction, $x=L-1$. As you cross the $x$ boundary, you have a discontinuity in the link variables, unless you choose
\be\label{eq:periodic}
	qa^2 B = \frac{2\pi n}{L}\ .
\ee
Of course, as pointed out in various references \cite{Bernard:1982yu, Lee:2005ds}, for modest volumes and lattice spacings used in simulations, this requires magnetic fields that are so large that we would expect nonlinearities to arise (and dominate) in the $B$-dependence of the masses, and possible distortions in the particles themselves. 

The best solution of course would be to use volumes large enough that these issues are irrelevant, however this can become rather expensive. What has been done is to ignore the na\"ive periodicity constraint in Eq.~(\ref{eq:periodic}). One would then use small fields that would not distort the particles and also ensure the linear relationship between the extracted mass in Eq.~(\ref{eq:mB}) and the magnetic field. One can also impose Dirichlet boundary conditions and place the source in the center of the lattice, and hope that this is sufficient to ensure the quarks will not feel the effects of the discontinuity.

The difficulty with this approach, however, is that there are significant finite volume effects in the results in the magnetic moments. Specifically, in the quenched calculation of Ref.~\cite{Lee:2005ds}, the authors see effects that are as large as 35\% for the lightest pion mass when going from a $16^3$ volume to a $24^3$ volume, with a lattice cutoff of $a^{-1}\approx 2\ {\rm GeV}$, and a pion mass of about 522 MeV. Since taking the pion mass closer to the physical point, finite volume errors become more substantial, we would like to reduce the finite volume effects coming from the background field as much as possible.

It was first pointed out in Ref.~\cite{Damgaard:1988hh} that the periodicity constraint shown in Eq.~(\ref{eq:periodic}) is not the best constraint possible. In fact, there is a more physically reasonable way to implement the background field such that we can reduce the minimally allowed magnetic field and still satisfy the continuity requirements.\footnote{We thank Taku Izubuchi for pointing this out to us.}

First we note that it is not the link that should to be continuous over a boundary, but instead we want the flux through the boundary to be continuous. This is what corresponds, in the continuum limit, to the magnetic field directly. Thus, we should be assured that the plaquette is continuous across the boundary. With the current implementation of the magnetic field, the forward plaquettes in the $x-y$ plane are given by:
\bea
	e^{-iqa^2 B} \ \ \text{for\ } x\ne L-1\ ,
	&\quad&
 	e^{iqa^2 B(L-1)} \ \ \text{for\ } x= L-1\ .
	\label{eq:plaquettes_boundaries}
\eea
There is no choice of the magnetic field which would allow the plaquette to be continuous across the boundary.

We can make the plaquettes continuous by ``patching'' the field: adding the following $x$-link modification with Eq.~(\ref{eq:y-link-mod}):
\be\label{eq:x-link-mod}
	A_\mu(L-1,y,z,t) = -a B L y\; \delta_{\mu x} \ \ \text{if\ } x=L-1  
	\ .
\ee
So now a plaquette on the $x$-boundary is given by $e^{-iqa^2B}$ for $x= L-1, y\ne L-1$. When $y=L-1$ we have $e^{iqa^2B(L^2-1)}$ along the $x$ boundary and for a continous boundary, we must have
\be\label{eq:patched_periodicity}
	q a^2 B = \frac{2\pi n}{L^2}\ .
\ee
Thus, we gain more than an order of magnitude for the minimally allowed magnetic field with this additional (and ``free'') modification. The question here is whether or not this is a sufficient amount to ensure the minimal $B$-field is small enough to not distort the baryons.

For $n=\pm1,\pm2$, we would assume it is. However, to have two different magnetic fields to extract the $\Delta^+$ magnetic moment, we must also use a third field with $n=\pm 4$, and one needs to worry whether or not this field is now large enough for distortions to appear. If it is, one would like to know whether or not it is safe to neglect the periodicity constraint as was done in previous studies if we perform this patching. We study this in the next section.

\section{Tests on quenched lattices}

For an initial test, we calculated the magnetic moment of the $\Delta^{++}$ on two quenched anistropic ensembles, with volumes $16^3\times64$ (generated by the JLab lattice group) and $24^3\times128$, with a spatial lattice spacing of 0.1 fm and an anisotropy of $a_s/a_t \approx 3$. The pion mass for these lattices is $\sim$750 MeV, so that we are assured to not have any finite volume effects coming from a small $m_\pi L$. Also, we use standard periodic boundary conditions in the spatial directions, and we place the source of the quark propagator in the center of the spatial lattice.

\begin{figure}[t]
\begin{center}
\includegraphics[width=4in]{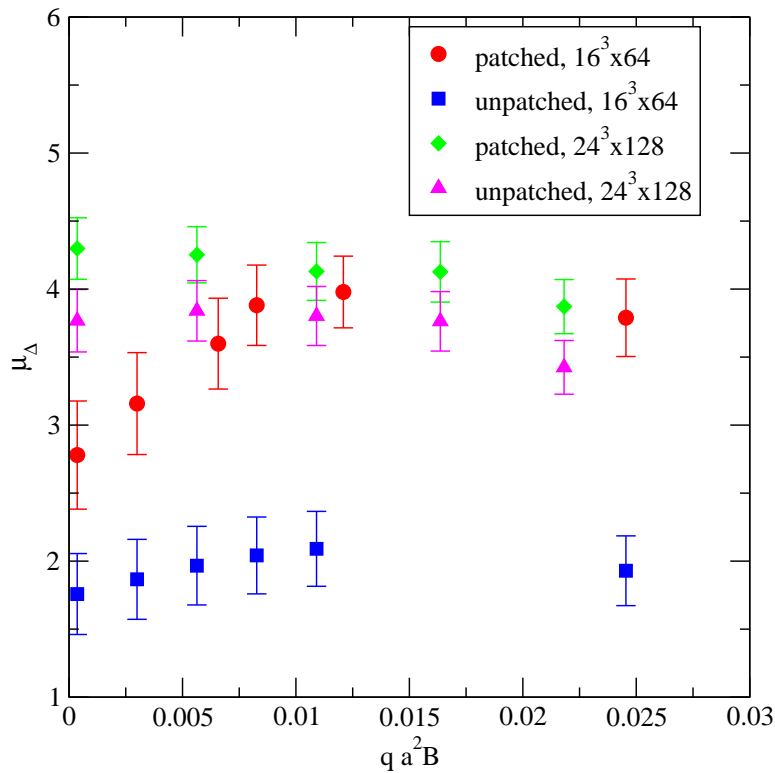}
\caption{Comparing the patched results to the unpatched results for the two quenched volumes studies. Notice the large finite volume effects on the small volume, either for all fields (unpatched), or merely for small fields (patched).}
\label{fig:p_vs_unp}
\end{center}
\end{figure}

We show in Fig.~\ref{fig:p_vs_unp} the results from this test. The two small volume results are in red (patched, so the minimum $q a^2 B$ that satisfies the periodicity constraint is 0.025) and in blue [unpatched, so $(qa^2B)_{\rm min} = 0.39$, too large to be simulated], while the two large volume results are in green [patched $(qa^2B)_{\rm min} = 0.011$], and purple [unpatched $(qa^2B)_{\rm min} = 0.26$, also too large to be simulated]. 

What we see is that the unpatched, small volume data disagrees by a factor of two from all three other data sets. Additionally, the patched small volume data tends to agree with the large volume data for larger $qa^2B$ (near the first periodic point), but is decreasing as one decreases $q a^2 B$, such that for the smallest $q a^2 B$ simulated, there is a noticeable discrepancy. In contrast, the large volume data (where the first two periodic points appear on the plot at $q a^2 B = 0.011,0.022$) is consistent for all values simulated. 

This indicates that even on the small volume (where simulations are cheaper), as long as one includes this additional modification, one can obtain trustworthy results, if the $B$-fields simulated are not too far from the minimum allowed $B$-field. Clearly while using smaller magnetic fields would ensure no distortion of the hadron under study, it creates an extremely large discontinuity at the boundary, which is seen very strongly in the figure. Similar finite-volume effects were seen in Ref.~\cite{Lee:2005ds}, where they used small magnetic fields and spatial Dirichlet boundary conditions, which was supposed to ensure that the hadron would not see the large discontinuity at the boundary. Obviously this choice of boundary conditions does not hide the discontinuity from the hadron for the small volume. Our study indicates that implementing the ``patched'' $B$-field will allow for a small enough discontinuity (in Fig.~\ref{fig:p_vs_unp}, $qa^2B\ge0.0075$ roughly), that one can trust the small volume data. Thus, we conclude that it is essential to include this modification to obtain reliable physical results.

\section{Dynamical results}\label{sec:dynamical}

Now we would like to present some initial (and preliminary) dynamical results for some magnetic moments using a background field technique. We note that these are the first dynamical results for magnetic moments using a background field. In this section, we will only use results that come from patched magnetic fields, thus saving us from significant finite volume effects that would come from a discontinuous magnetic field on the boundaries. This of course does not mean we need not worry about finite volume errors, as there may still be some coming from the pion mass. 

We use dynamical anisotropic lattices with 2+1 flavors of Stout-smeared Clover fermions \cite{Edwards:2008ja,PeardonLat}, on two volumes and, a single lattice spacing. For this initial study, we have a single light quark mass on both volumes, with 147 configurations used for the light quark propagators on the large $24^3\times128$ volume and 39 on the small $16^3\times128$ volume. We also generated 120 strange quark propagators on the large volume to obtain the $\Omega^-$ magnetic moment. On these lattices, a bare light quark mass parameter of $-0.0840$ corresponds to a pion mass of 366 MeV, and $-0.0743$ corresponds roughly to the physical strange quark mass. Both sets of configurations have an inverse spatial lattice spacing of 1.8 GeV and an anisotropy of 3. More complete information on these configurations, including how the tuning of the parameters have been done, can be found in \cite{Edwards:2008ja,PeardonLat}.

In all cases we used two magnetic fields for the simulations. These correspond to
\be\label{eq:Bfield_dyn}
	q a^2 B = 
	\pm\frac{\pi}{L^2}
	\ ,\ \pm\frac{2\pi}{L^2}\ ,
\ee
so the first point does not satisfy the periodicity constraint. We expect the errors entering here due to this to be negligible as we showed in the previous section. Doing so allows us to not worry that the magnetic field will distort our baryons, but so that we can have two fields with which to simulate the $\Delta^{++,-},\Omega^{-}$ and one for the $\Delta^{+,0}$.\footnote{In an upcoming publication, we will have results for three input magnetic fields.} We calculate all four spin projections for the baryons, as well as using both positive and negative magnetic fields, and we average over all of these to reduce our errors. In these simulations, the relevant baryon masses are $m_\Delta = 1.408$ GeV and $m_{\Omega^-} = 1.65$ GeV. 

\begin{figure}[t]
\begin{center}
\includegraphics[width=5.9in]{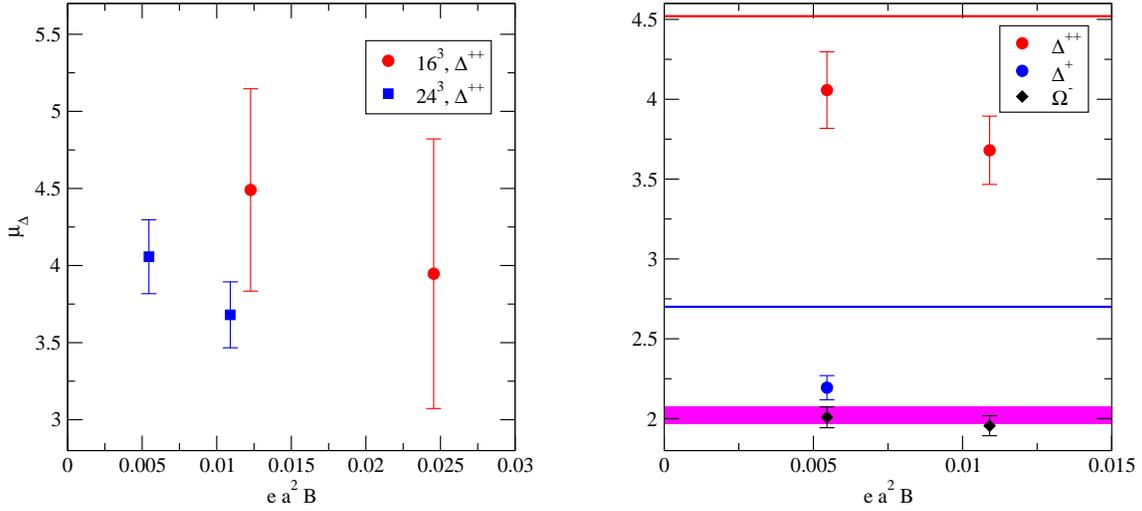}
\caption{The left plot is a comparison of the $\Delta^{++}$ magnetic moment on the large and small volumes. The right plot is a comparison of $\mu_\Delta$ (blue and red) and $|\mu_{\Omega^-}|$ (the black points are lattice data, the magenta line is experiment) with experiment.}
\label{fig:deltamm}
\end{center}
\end{figure}

We show two plots in Fig.~\ref{fig:deltamm} for the magnetic moments, all of which are in units of the physical nuclear magneton $\mu_N$. The first compares the $\Delta^{++}$ magnetic moment on the two volumes simulated. The fact that the results are consistent may seem of limited value due to the small number of configurations used on the small volume. However, given the factor of two discrepancy in the quenched test between the small and large volumes (in the unpatched case), this agreement shows that clearly, even for this smaller pion mass, the small volume result will not differ greatly from the large volume result. In an upcoming publication, we will use this fact to simulate heavier pion masses on the smaller volume, where finite volume effects coming from the pion mass should not be terribly large.

The second plot in Fig.~\ref{fig:deltamm} compares the large volume $\Delta^{++}$, $\Delta^+$, and (the absolute value of the) $\Omega^-$ magnetic moments to the central experimental values (note the errors on the $\Delta$ experimental numbers would fill the plot). While the central values differ for the $\Delta$'s, the results are in complete agreement considering not just the large experimental errors, but the fact that we are working at an unphysical pion mass. Clearly simulating at multiple pion masses is necessary for a chiral extrapolation, which we are currently working on. 

As for the $\Omega^-$, the strange quark mass is close to its physical value (as we can see by the fact that the $\Omega^-$ mass is close to the observed value), so we expect the result to match more closely to the experimental value, and as we can see, it does for both values of the $B$-field simulated. This agreement is expected, as quantities involving the $\Omega^-$ should have little dependence on the light sea quark mass. Additionally, we see that the errors associated with the experimental value are comparable to the statistical lattice errors here.

\section{Conclusion}\label{sec:conc}

We conclude by first stressing the primary result of this work: In order to minimize finite volume effects coming from discontinuities associated with the background field, one should always implement a patched background field, where in addition to a traditional implementation of the background field, boundary links are modified to ensure the plaquettes can be continuous over the boundary (this is also done in Ref.~\cite{Detmold:2008xk} for polarizabilities). When this is done, however, we find that we do not have to satisfy this periodicity, so long as we remain close to the first periodic point. Otherwise before one worries about finite volume effects coming from small pion masses, there will be significant finite volume effects from the background field. By implementing this patched condition, we allow ourselves to reliably extract magnetic moments even from smaller volumes, where otherwise the data would suffer greatly from finite volume effects.

We also showed the first results for the $\Delta$ and $\Omega^-$ magnetic moments on dynamical 2+1-flavor lattices, which are consistent (given the pion mass used) with experimental values that have been measured. Presently, the accuracy obtained in the lattice result for the $\Omega^-$ magnetic dipole moment is comparable with the experimental accuracy. In the near future, we will be adding more statistics as well as understanding the quark mass dependence of these quantities.

\bigskip

We would like to thank NERSC and USQCD for the computing resources used to carry out this study, as well as the Jefferson Lab Lattice group for the anisotropic clover lattices. Additional analysis was done on the cyclades computer cluster at WM. This work was partially supported by the US Department of Energy, under contract nos. DE-AC05-06OR23177 (JSA), DE-FG02-07ER41527, and DE-FG02-04ER41302; and by the Jeffress Memorial Trust, grant J-813.

\bibliography{refs}


\end{document}